\documentclass[showpacs,10pt,twocolumn,prl]{revtex4-1}
\usepackage{amsmath}
\usepackage{amssymb}
\usepackage{graphics}
\usepackage{epsfig}
\usepackage{CJK}

\begin{document}


\title{Superconductivity in K$_{x}$Fe$_{2-y}$Se$_{2-z}$S$_{z}$ (0 $\leqslant
$ z $\leqslant $ 2) and Lattice Distortion}
\author{Hechang Lei,$^{1}$ Milinda Abeykoon,$^{1}$ Emil S. Bozin,$^{1}$
Kefeng Wang,$^{1}$ J. B. Warren,$^{2}$ and C. Petrovic$^{1}$}
\affiliation{$^{1}$Condensed Matter Physics and Materials Science Department, Brookhaven
National Laboratory, Upton, New York 11973, USA\\
$^{2}$Instrumentation Division, Brookhaven National Laboratory, Upton, New
York 11973, USA}
\date{\today}

\begin{abstract}
We report structurally tuned superconductivity in K$_{x}$Fe$_{2-y}$Se$_{2-z}$%
S$_{z}$ (0 $\leqslant $ z $\leqslant $ 2) phase diagram. Superconducting $%
T_{c}$ is suppressed as S is incorporated into the lattice, eventually
vanishing at 80\% of S. The magnetic and conductivity properties
can be related to stoichiometry on poorly occupied Fe1
site and the local environment of nearly fully occupied Fe2
site. The decreasing $T_{c}$ coincides with the increasing Fe1
occupancy and the overall increase in Fe stoichiometry from $z=0$ to $z=2$.
Our results indicate that the irregularity of Fe2-Se/S tetrahedron is an
important controlling parameter that can be used to tune the ground state in
the new superconductor family.
\end{abstract}

\pacs{74.62.Bf, 74.10.+v, 74.20.Mn, 74.70.Dd}

\maketitle

The discovery of superconductivity in K$_{x}$Fe$_{2-y}$Se$_{2}$ (AFeCh-122
type) with $T_{c}$\ $\approx $ 31 K has triggered a renewed interest in the
field of iron-based superconductors \cite{Guo}. The crystal structure
contains alkali metals and FeSe layers alternatively stacked along the c
axis. When compared to other iron-based superconductors, AFeCh-122 show some
distinctive features. Band structure calculations \cite{Shein} and angle
resolved photoemission spectroscopy (ARPES) measurements \cite{Zhang Y}
indicate that the hole pockets are absent in AFeCh-122 system. This is
different from other iron-based superconductors where both electron and hole
Fermi surfaces (FSs) are present \cite{Ding H1}, suggesting that
unconventional pairing via FS nesting \cite{Mazin},\cite{Kuroki} might not
be suitable for these materials. Experiments observed short and possibly
long range magnetic order and its coexistence with superconductivity \cite%
{Torchetti},\cite{Pomjakushin},\cite{Pomjakushin2}. This is quite different
from previous iron-based superconductors where superconductivity emerges
when the spin density wave (SDW) state is suppressed \cite{Cruz}. Finally,
the role of the ordered Fe vacancy and its relations to the magnetic order,
metallic states and superconductivity remains unresolved.

Isovalent substitution is an effective way to study the relationship between
structural parameters and physical properties. It is similar to pressure
effects, because it should not change the carrier density. Unlike in the
cuprates, superconductivity has been induced or enhanced by isovalent doping
in iron-based superconductors, such as BaFe$_{2}$(As$_{1-x}$P$_{x}$)$_{2}$
and FeTe$_{1-x}$Se$_{x}$ \cite{Jiang S},\cite{Yeh}.\ Recently, it was
discovered that K$_{x}$Fe$_{2-y}$S$_{2}$, isostructural to K$_{x}$Fe$_{2-y}$%
Se$_{2}$, exhibits a spin glass (SG) semiconductor ground state without
superconducting transition \cite{Lei HC2}. Here, we report the evolution of
physical properties and structural parameters of K$_{x}$Fe$_{2-y}$Se$_{2-z}$S%
$_{z}$ ($0\leqslant z\leqslant 2$) single crystals. The results indicate
that suppression of superconducting $T_{c}$ coincides with the increasing
occupancy of Fe on both Fe1 and Fe2 sites with S substitution, in contrast
to (Tl,K)$_{x}$Fe$_{2-y}$Se$_{2}$ \cite{Fang MH}. It is found that the
regularity of Fe2-Ch (Ch = Se, S) tetrahedron is an important controlling
parameter for electronic and magnetic properties and has a major influence
on electron pairing, i.e. on $T_{c}$.

Single crystals of K$_{x}$Fe$_{2-y}$Se$_{2-z}$S$_{z}$ were grown as
described previously \cite{Lei HC2},\cite{Lei HC}. Powder X-ray diffraction
(XRD) data were collected at 300 K using 0.3184 \AA\ wavelength radiation
(38.94 keV) at X7B beamline of the National Synchrotron Light Source.\
Refinements of the XRD data were performed using General Structure Analysis
System (GSAS) \cite{Larson},\cite{Toby}. The average stoichiometry was
determined by energy-dispersive x-ray spectroscopy (EDX) in a JEOL JSM-6500
scanning electron microscope. Electrical transport and magnetization
measurements were carried out in Quantum Design PPMS-9 and MPMS-XL5.

\begin{figure}[tbp]
\centerline{\includegraphics[scale=0.42]{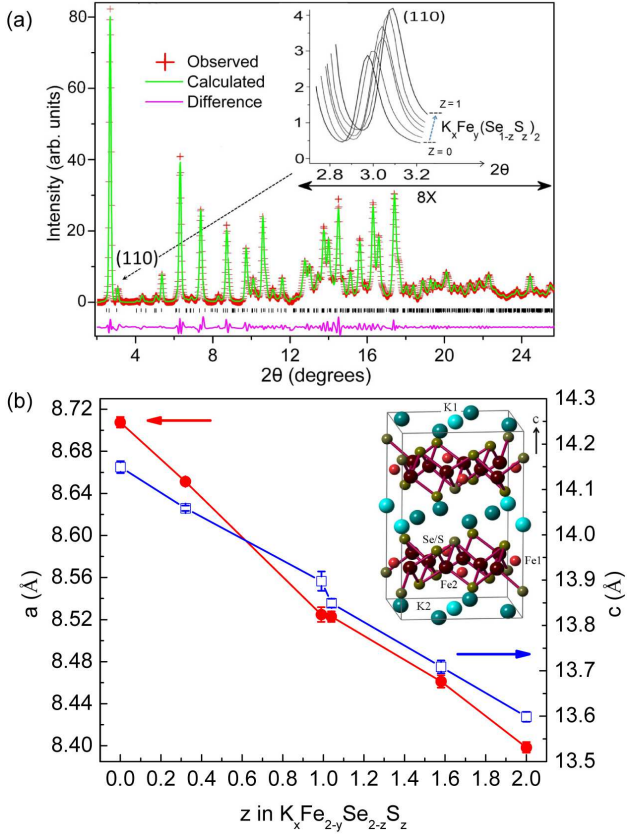}} \vspace*{-0.3cm}
\caption{(a) Powder XRD patterns of K$_{x}$Fe$_{2-y}$Se$_{2-z}$S$_{z}$ at
300 K and fit using I4/m space group. Inset: (110) peak position with
different S content. (b) Unit cell parameters as a function of S
substitution. Inset: Crystal structure of K$_{x}$Fe$_{2-y}$Se$_{2-z}$S$_{z}$
in I4/m unit cell with vacant Fe1 sites marked light red and Fe2 sites
marked dark purple. }
\end{figure}

XRD patterns of all K$_{x}$Fe$_{2-y}$Se$_{2-z}$S$_{z}$ show (110) and other
superlattice reflections associated with I4/m symmetry that incorporates Fe
ordered vacancy site (Fig. 1(a)). The structure is shown in the inset of
Fig. 1(b).\ The peak position of (110) shifts to higher angle with
increasing S content, which indicates that lattice contracts gradually with
S doping (Fig. 1(b) and Table 1), due to the smaller ionic size of S$^{2-}$
than Se$^{2-}$. The trend of lattice contraction approximately follows\ the
Vegard's law. The sum of Se and S stoichiometry in grown crystals is near
2.0, whereas both potassium and iron deviate from full occupancy (Table 1)
\cite{Lei HC2},\cite{Lei HC}. This is commonly found in all AFeCh-122
compounds \cite{Guo},\cite{Lei HC2},\cite{Wang DM}. It should be noted that
the Fe content increases gradually with S doping, while the K content does
not change monotonically. In addition, doping S into Se site does not lead
to random Fe occupation of Fe1 and Fe2 sites. The refinements show that the
Fe1 site is poorly occupied, whereas the Fe2 site is almost fully occupied.

\begin{table*}[tbp] \centering%
\caption{Structural, magnetic and transport properties of
K$_{x}$Fe$_{2-y}$Se$_{2-z}$S$_{z}$. $T_{f}$ is determined from maximum position of $\chi_{ab}(T)$ in zero-field-cooling curve. $T_{\rho _{max}}$, $T_{c, onset}$ and $T_{c, 0}$ are obtained from
resistivity data. Error bars for $T_{c}$, $T_{f}$, and $T_{\rho _{max}}$ reflect standard deviation from the average temperatures measured on several crystals grown in one batch.}%
\begin{tabular}{ccccccc}
\hline\hline
K$_{x}$Fe$_{2-y}$Se$_{2-z}$S$_{z}$ & a (\AA ) & c (\AA ) & $T_{c,onset}$ (K)
& $T_{c,0}$ (K) & $T_{f}$ (K) & $T_{\rho _{max}}$ (K) \\ \hline
K$_{0.64(4)}$Fe$_{1.44(4)}$Se$_{2.00(0)}$ & 3.911(2) & 14.075(3) & 33(1) &
31(1) &  & 132(6) \\
K$_{0.73(3)}$Fe$_{1.44(3)}$Se$_{1.68(6)}$S$_{0.32(5)}$ & 3.847(2) & 14.046(4)
& 33(1) & 31(1) &  & 129(4) \\
K$_{0.70(7)}$Fe$_{1.55(7)}$Se$_{1.01(2)}$S$_{0.99(2)}$ & 3.805(2) & 13.903(6)
& 24(2) & 21(1) &  & 203(20) \\
K$_{0.76(5)}$Fe$_{1.61(5)}$Se$_{0.96(4)}$S$_{1.04(5)}$ & 3.797(2) & 13.859(3)
& 18(1) & 16(2) &  & 138(5) \\
K$_{0.80(8)}$Fe$_{1.64(8)}$Se$_{0.42(5)}$S$_{1.58(0.05)}$ & 3.781(2) &
13.707(2) &  &  & 17(1) & 79(14) \\
K$_{0.80(1)}$Fe$_{1.72(1)}$S$_{2.00(1)}$ & 3.753(2) & 13.569(3) &  &  & 32(1)
&  \\ \hline\hline
\end{tabular}%
\label{TableKey}%
\end{table*}%
\begin{figure}[tbp]
\centerline{\includegraphics[scale=0.42]{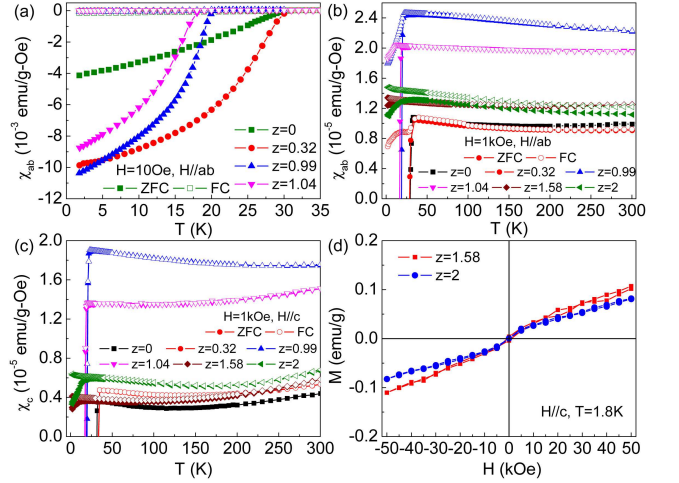}} \vspace*{-0.3cm}
\caption{(a) dc magnetic susceptibility of superconducting K$_{x}$Fe$_{2-y}$%
Se$_{2-z}$S$_{z}$ samples below 35 K at H = 10 Oe for H$\Vert $ab in ZFC and
FC. (b) and (c) Temperature dependence of $\protect\chi $(T) of K$_{x}$Fe$%
_{2-y}$Se$_{2-z}$S$_{z}$ at H = 1 kOe for H$\Vert $ab and H$\Vert $c in ZFC
and FC, respectively. (d) M-H loops of non-superconducting samples at 1.8 K
for H$\Vert $c.}
\end{figure}

The superconductivity of K$_{x}$Fe$_{2-y}$Se$_{2-z}$S$_{z}$ single crystals
with z $<$ 1.58 is confirmed by the magnetization measurement (Fig. 2(a)).
Zero-field-cooling (ZFC) superconducting transitions shift to low
temperature with higher S content. Field-cooling (FC) susceptibilities are
very small (Fig. 2(a)), implying possible strong vortex pinning. The most
interesting feature in temperature dependence of magnetization is the
absence of Curie-Weiss law for all samples above 50 K as shown in Fig. 2(b)
and (c) for H$\parallel $ab and H$\parallel $c, respectively. Magnetic
susceptibilities are weakly temperature dependent with no significant
anomalies above 50 K. This might suggest the presence of low dimensional
short-range magnetic correlations and/or a long range antiferromagnetic
(AFM) order above 300 K. This has been observed in K$_{x}$Fe$_{2-y}$Se$_{2}$
crystals with different K and Fe sites occupancies \cite{Torchetti},\cite%
{Pomjakushin2}. Therefore the magnetic interactions above 50 K are similar
in the whole alloy series. On the other hand, non-superconducting samples
show bifurcation between the ZFC and FC curves for both field orientations.
The irreversible behavior suggests spin glass (SG) transition at low
temperatures near $z=2$ \cite{Lei HC2}. The spin glass freezing temperatures
$T_{f}$ for non-superconducting samples are determined from maximum position
of $\chi _{ab}(T)$ in ZFC curves and listed in Table 1. The S-shape M(H)
curves (Fig. 2(d)) at 1.8 K for $z=1.58$ and $z=2$ are similar \cite{Lei HC2}%
,\cite{Ying JJ2}. Hence, the SG state most likely persists in the 1.58 $%
\leqslant $ z $\leqslant $ 2 range.
\begin{figure}[tbp]
\centerline{\includegraphics[scale=0.45]{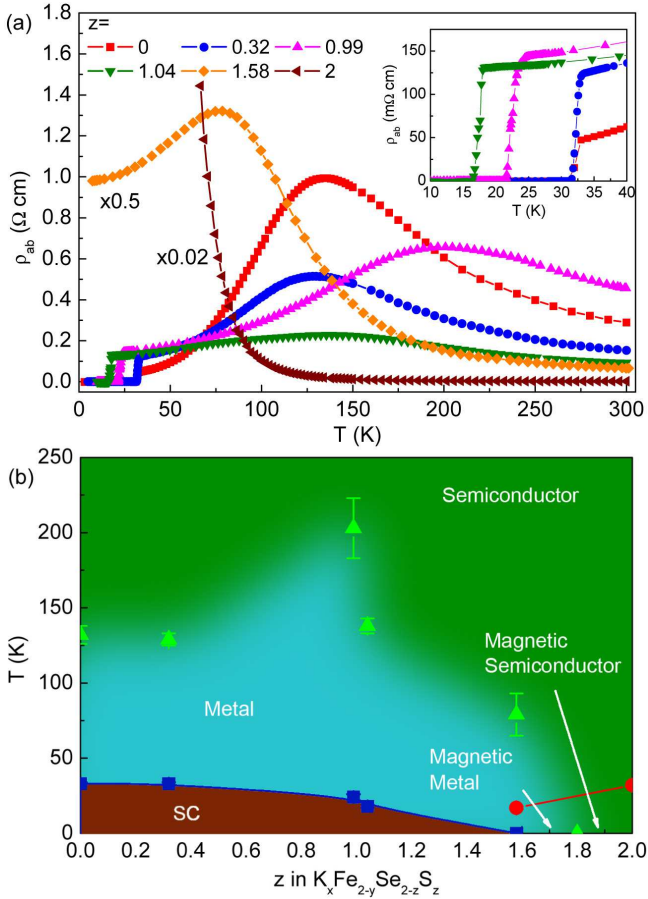}} \vspace*{-0.3cm}
\caption{(a) Temperature dependence of the in-plane resistivity $\protect%
\rho _{ab}(T)$ of the K$_{x}$Fe$_{2-y}$Se$_{2-z}$S$_{z}$ single crystals in
zero field. Inset: temperature dependence of $\protect\rho _{ab}(T)$ below
40 K for $0\leqslant z\leqslant 1.2$. (b) Magnetic and superconducting phase
diagram of K$_{x}$Fe$_{2-y}$Se$_{2-z}$S$_{z}$. Green, blue and orange colors
show semiconducting, magnetic and superconducting regions, respectively. Red
symbols denote spin glass transitions.}
\end{figure}

Besides K$_{x}$Fe$_{2-y}$S$_{2}$, all other crystals show metallic behavior
below a resistivity maximum $\rho _{\max }$ above superconducting transition
(Fig. 3(a)). It should be noted that the temperature of $\rho _{\max }$ is
not monotonic with the doping level of S (z), implying that the crossover
may be influenced by the amount of both K and Fe deficiencies (Table 1). On
the other hand, the crystal with $z=2$ is semiconducting even if the Fe
deficiency is smaller than in other samples. In contrast to $\rho _{\max }$,
with the increase in S, the $T_{c}$ is monotonically suppressed to lower
temperature and can not be observed above 2 K for z $\geqslant $ 1.58 (Table
1). We present the magnetic and superconducting phase diagram of K$_{x}$Fe$%
_{2-y}$Se$_{2-z}$S$_{z}$ in Fig. 3(b). Semiconductor-metal crossover can be
traced for $0\leqslant z\leqslant 1.58$ at high temperature. In this region,
K$_{x}$Fe$_{2-y}$Se$_{2-z}$S$_{z}$ is a superconducting metal at low
temperature. For $z=1.58$, $\rho $(T) is metallic with no superconducting $%
T_{c}$ down to 2 K. For $1.58\leqslant z\leqslant 2$, we observe a drop in $%
\chi -T$ curves that could be related to a SG transition. For $z=2$, the K$%
_{x}$Fe$_{2-y}$Se$_{2-z}$S$_{z}$ becomes a small gap semiconductor with no
metallic crossover and with a SG transition below 32 K.

The gradual changes of $T_{c}$ are difficult to explain by the slight
variation of Fe and K contents, because the $T_{c}$ variation found with S
doping is rather different from $T_{c}$ variations due to K and Fe
differences \cite{Fang MH},\cite{Wang DM},\cite{Yan YJ}. It was shown that
the superconductivity appears with higher Fe content when K content (x) $<$
0.85 \cite{Wang DM},\cite{Yan YJ}. As opposed to this trend, the K$_{x}$Fe$%
_{2-y}$Se$_{2-z}$S$_{z}$ crystals with larger z values have higher Fe
content but lower $T_{c}$.

\begin{figure}[tbp]
\centerline{\includegraphics[scale=0.4]{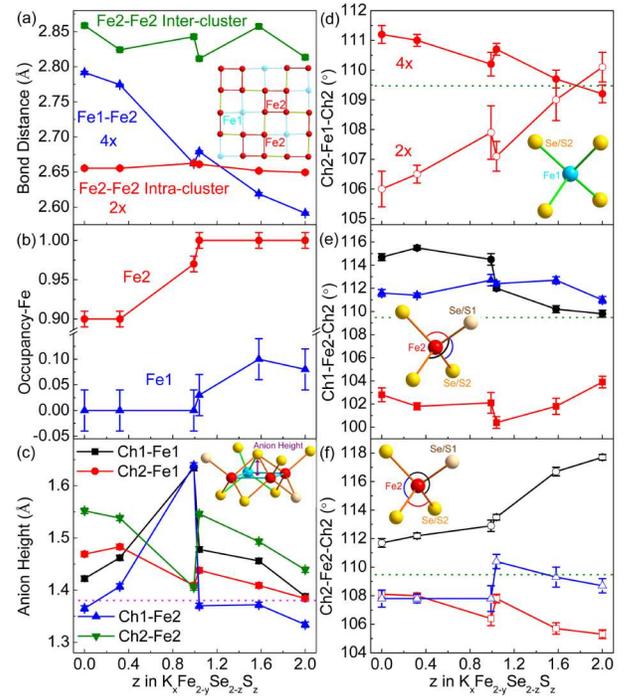}} \vspace*{-0.3cm}
\caption{(a) Bond lengths between Fe1(2) and Fe1(2). Inset shows top view of
Fe layer. (b) Occupancy of Fe1 and Fe2. (c) Anion height. Inset shows side
view of Fe-Ch sheet. (d) Ch2-Fe1-Ch2, (e) Ch1-Fe2-Ch2, and (f) Ch2-Fe2-Ch2
bond angles of K$_{x}$Fe$_{2-y}$Se$_{2-z}$S$_{z}$ as a function of S
substitution, $x$. Insets of (d)-(f) show Fe-Ch tetrahedron for Fe1-Ch1 and
Fe2-Ch1(2). The dotted pink line in (c) shows the optimal bond lengths for $%
T_{c}$. The dotted green lines in (d)-(f) indicate the optimal angle for $%
T_{c}$.}
\end{figure}

Why is K$_{x}$Fe$_{2-y}$Se$_{2}$ a superconductor whereas K$_{x}$Fe$_{2-y}$S$%
_{2}$ is a semiconductor even though S is an isovalent substitution, similar
to FeTe$_{1-x}$Se$_{x}$ \cite{Yeh}? This testifies that $T_{c}$ is not only
governed by K/Fe stoichiometry or vacancies. Below we show that
superconductivity is in fact tuned by subtle structural effects induced by
stoichiometry variation. The local environment of Fe profoundly changes with
S, inducing the evolution of band structure and changes of physical
properties. In the K$_{x}$Fe$_{2-y}$Se$_{2-z}$S$_{z}$ crystal structure, Fe
atoms have block-like distribution where every four Fe2 form a square around
Se atom, making a cluster distinct from Fe1 site with low occupancy \cite%
{Yin WG},\cite{Bao W}. Therefore there are Fe1-Fe2 distances as well as
intra- and inter-cluster Fe2-Fe2 distances. The three types of distances are
shown in Fig. 4(a). All cluster distances are unchanged with S doping
whereas the Fe1-Fe2 distances decrease significantly (Fig. 4(a)).\ Similar
magnetization behavior above 50 K and rather different superconducting $T_{c}
$'s as S content varies from 0 to 2 coincides with the nearly unchanged
Fe2-Fe2 bond lengths. This shows that superconductivity is insensitive to
the size of Fe2-Fe2 clusters whereas the unchanged high temperature
magnetism could be related to the unchanged Fe2-Fe2 bond lengths. On the
other hand, the SG behavior arising at the low temperature for
non-superconducting samples can be explained by the non zero random
occupancy of Fe1 (vacancy) site for higher S content (Fig. 4(b)), randomly
changing the inter-cluster exchange interactions \cite{Oledzka}.

According to the empirical rule proposed by Mizuguchi et al., the critical
temperature is closely correlated with the anion height between Fe and Pn
(Ch) layers and there is an optimal distance ($\sim $ 1.38 $\mathring{A}$)
with a maximum transition temperature $T_{c}$ $\sim $ 55 K \cite{Mizuguchi3}%
. This is invalid in K$_{x}$Fe$_{2-y}$Se$_{2-z}$S$_{z}$ materials, where
there are two Fe and two Ch sites and four Fe-Ch heights. This is because
there is no monotonic decrease as $T_{c}$ is tuned to 0 (Fig. 4(c)), whereas
both Se and S end members have rather similar anion heights.

The bond angle $\alpha $ between Pn(Ch)-Fe-Pn(Ch) is another important
factor since $T_{c}$ in iron pnictides is optimized when Fe-Pn (Ch)
tetrahedron is regular ($\alpha $ = 109.47$^{\circ }$) \cite{Lee}. In K$_{x}$%
Fe$_{2-y}$Se$_{2-z}$S$_{z}$, the Ch2-Fe1-Ch2 angle changes towards the
optimal value with increasing S (Fig. 5(d)), but the local environment of
Fe2 exhibits inverse trend. Among six angles in Fe2-Ch1(2) tetrahedron,
three (Ch1-Fe2-Ch2) are nearly unchanged with S doping (Fig. 4(e)). The
other three (Ch1-Fe2-Ch1) change significantly (maximum 6$^{\circ }$) and
deviate from optimal value from Se rich to S rich side (Fig. 4(f)). Hence,
the increasing distortion of Fe2-Ch tetrahedron with S doping is closely
correlated to the suppression of $T_{c}$. This distortion may lead to
carrier localization, decreasing the density of states at the FS. Therefore,
the regularity of Fe2-Ch1(2) tetrahedron is the key structural factor in the
formation of the metallic states in K$_{x}$Fe$_{2-y}$Se$_{2-z}$S$_{z}$, and
consequently, the $T_{c}$. The dispersion of tetrahedral angles from 109.47$%
^{\circ }$ increases with S content increase.

We have demonstrated structure tuning of superconductivity in K$_{x}$Fe$%
_{2-y}$Se$_{2-z}$S$_{z}$ single crystals. For low S doping, the
superconducting $T_{c}$ is nearly the same as in the pure material. With the
increase of S, above $z=1$ the $T_{c}$ is gradually suppressed finally
vanishing at 80\% of S substitution. Conductivity and magnetic properties
coincide with stoichiometry changes on Fe1 site and with particular changes
of the local environment of Fe2 site. Nearly unchanged Fe2-Fe2 bond lengths
result in the similar magnetic behavior. The suppression of
superconductivity with S doping could be traced to the increasing
irregularity of Fe2-Ch tetrahedron and to increasing occupancy of Fe1 site.
These ultimately destroy superconductivity and bring about glassy magnetic
order and semiconducting ground state.

We thank Jonathan Hanson for help with XRD experiment. Work at Brookhaven is
supported by the U.S. DOE under Contract No. DE-AC02-98CH10886 and in part
by the Center for Emergent Superconductivity, an Energy Frontier Research
Center funded by the U.S. DOE, Office for Basic Energy Science.

\end{document}